# A quantum-inspired predictor of Parkinson's disease built on a diverse, multimodal dataset

Diya Vatsavai[1], Anya Iyer[2], Ashwin A. Nair[3]*

[1] Valley Christian High School, San Jose, California, United States of America

diya.vatsavai08@gmail.com

[2] Dougherty Valley High School, San Ramon, California, United States of America.

anya.iyer@gmail.com

[3] UC Davis Graduate School of Management, Davis, California, United States of America

aaravind@ucdavis.edu*

*Corresponding Author

**Abstract**

Parkinson's disease, the fastest-growing neurodegenerative disorder globally, has seen a 50% increase in cases within just two years. As speech, memory, and motor symptoms worsen over time, early diagnosis is crucial for preserving patients' quality of life. While machine-learning-based detection has shown promise, relying on a single feature for classification can be error-prone due to the variability of symptoms between patients. To address this limitation we utilized the mPower database, which includes 150,000 samples across four key biomarkers: voice, gait, tapping, and demographic data. From these measurements, we extracted 64 features and trained a baseline Random Forest model to select the features above the 80th percentile. For classification, we designed a simulatable quantum support vector machine (qSVM) that detects high-dimensional patterns, leveraging recent advancements in quantum machine learning. With a novel, simulatable architecture that can be run on standard hardware rather than resource-intensive quantum computers, our model achieves an accuracy of 90% and an AUC of 0.98—surpassing benchmark models. By utilizing an innovative classification framework built on a diverse set of features, our model offers a pathway for accessible global Parkinson's screening.

**Introduction**

Parkinson's Disease (PD) gradually damages brain cells, often leading to motor and nonmotor impairments [1]. PD is characterized by the degeneration of nerve cells in the brain's substantia nigra, where dopamine production occurs. As these cells lose their ability to produce dopamine, the body faces a deficiency in essential neurotransmitters. This dopamine shortage triggers a wide range of progressive symptoms, such as bradykinesia, hypokinetic dysarthria, resting tremor, and muscular rigidity [2]. Over the past 25 years, PD prevalence has doubled, and deaths linked to PD have increased by over 100% since 2000 [3]. These trends highlight an urgent need for effective detection and intervention methods.

Current detection methods, like DAT and PET scans, are not PD-specific, often leading to imprecise results [4]. These diagnostic approaches also require expensive, specialized equipment and medical expertise, costing the U.S. around $14 billion annually [5]. As the fastest-growing neurological condition globally, PD urgently calls for a low-cost, high-accuracy screening solution [6].

Despite recent validation of clinical diagnostic criteria, PD diagnosis remains challenging because PD symptoms overlap with those of other neurodegenerative diseases and aging. Diagnoses based on a single or limited set of clinical features often lead to inconclusive results. Research on specific biomarkers has helped improve PD detection, and in this study, we focus on vocal biomarkers, gait indicators, demographic data, and tapping metrics. Studies show that voice serves as a useful PD indicator. For instance, Ma et al.'s study, found that 70% to 90% of PD patients experience varying degrees of vocal impairment [7]. Voice changes, which often appear alongside tremor, emerge as early indicators of PD, with up to 78% of early-stage PD patients showing specific vocal changes [8]. Common vocal impairments associated with PD include pitch variations, decreased volume, unclear articulation, and an unstable voice. Given the diagnostic potential of speech impairments, we focus on analyzing specific vocal features associated with these changes, including volume, pitch, jitter, shimmer, and breathiness.

Gait serves as another characteristic marker of PD. Bradykinesia disrupts gait, causing both episodic and continuous disturbances [9]. Episodic disturbances include start hesitation and freezing of gait, while continuous disturbances reflect inconsistent walking patterns. These gait impairments significantly impact PD patients, with studies showing that 45%-68% experience falls annually, and 50%-86% fall recurrently [10]. Although gait disturbances aid in diagnosis, they also represent one of PD's most physically deteriorating symptoms, leading to severe injuries and heightening patients' fears of daily activities. Demographic data, particularly age, plays a well-established role in PD detection. The Parkinson's Foundation reports a sharp rise in PD incidence among individuals aged 65 and older, indicating that demographic factors could provide an early diagnostic window.

Finger-tapping data, which reflects bradykinesia, has proven useful yet controversial for PD diagnosis. Bradykinesia, a primary motor symptom of PD, manifests as slowness and halting in movement [11]. However, clinical evaluations of bradykinesia often rely on visual judgment [12], which varies widely between clinicians. The limited number of published studies on finger-tapping also lack robust interrater reliability estimates, emphasizing the lack of standardization. Nonetheless, some studies indicate that finger-to-thumb tapping tests correlate with lower overall UPDRS (Unified Parkinson's Disease Rating Scale) scores in PD patients [13]. Although the UPDRS offers a standardized method for assessing PD motor severity, it cannot fully address the issues of subjectivity and interrater variability.

Recent studies have explored the potential of machine learning algorithms to predict the occurrence of PD. However, most studies include a single feature for the basis of prediction as opposed to multiple features, resulting in performance metrics ranging from 60% to 85%, which are generally unsuitable for clinical applications [14, 15, 16, 17]. When it comes to diagnosis based on audio, convolutional neural networks (CNNs) have been applied to PD voice recordings, which are converted into spectrogram images for model classification. Many of these single-feature studies are built on private, homogenous datasets with minimal data points and unbalanced samples [18]. This can severely affect the reproducibility of the study as well as model generalizability due to lack of data. Finally, many studies involving voice use data made up of participants speaking a specific language, inflicting bias in classification [19].

To overcome the discussed limitations, our study's objective is to provide a multimodal diagnostic framework based on vocal data, gait tracking, tapping, and demographic information for comprehensive prediction, while removing the human subjectivity that arises with visual analysis of finger-tapping, by using automated detection. Our analysis is based on a diverse, publicly available dataset of over 150,000 samples, ensuring robustness while mitigating geographic bias. Furthermore, our vocal data consists of the universal syllable "ahh" for 10 seconds, free of language or accent bias. We utilize a custom quantum-assisted Support Vector Machine (qSVM) classifier exhibiting high performance even in classical simulations, removing the dependence on computationally expensive quantum hardware. With this model, we outperform standard machine learning techniques, state-of-the-art deep learning approaches, and commonly offered qSVM architectures with an accuracy of 90% and an ROC/AUC score of 0.98, all built on a diverse dataset representing participants with varied gender, age, racial and educational backgrounds, underscoring its potential for clinical applications globally.

**Results**

*Data Description:*

We utilized the mPower public research portal, which contains measurements from over 6,000 participants, both healthy and those affected by Parkinson's [20]. Certified researchers can access this publicly available dataset under protected access. The data includes common Parkinson's disease biomarkers: demographic information, such as age, gender, and smoking history, as well as voice recordings, tapping measurements, and gait tracking, all recorded through a smartphone app. Although many participants completed the same measurement multiple times, we randomly selected one trial per activity per participant for the final dataframe. This process helped us reduce any bias favoring participants who completed multiple trials. For model training and testing, we focused on 194 participants who completed the voice, gait, and tapping tests. This subset stands out for its diversity, including male and female participants who identified as Caucasian, African American, Hispanic, East Asian, South Asian, and mixed race. The subset also represents a range of educational backgrounds, with 35% of participants not holding a four-year college degree. We divided the dataset into 164 training samples and 30 test samples, maintaining an equal balance between Parkinson's-affected and healthy individuals.

*Feature Selection:*

Using this dataset, we extracted 64 voice, gait, tapping, and demographic features for each of the 194 participants, balanced between healthy individuals and those with Parkinson's. Additional details on feature extraction appear in the methods section. We normalized the dataframe using Scikit Learn's StandardScaler to ensure a consistent magnitude for each feature. Then, we trained a baseline Random Forest model to identify the top-performing features for the final qSVM model, selecting features with importance values above the 80th percentile [21] (Figure 1).

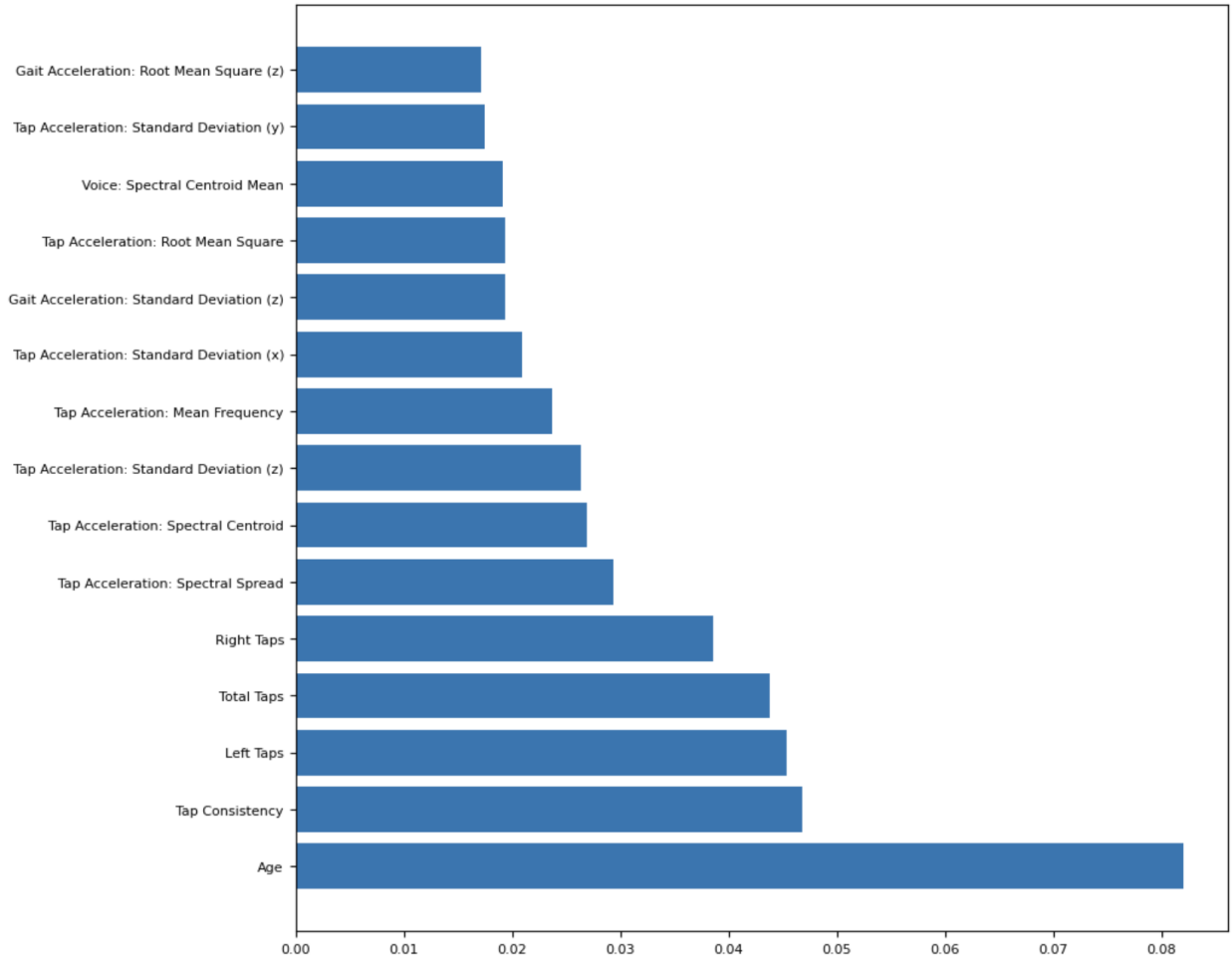

Figure 1: Feature importance values for features above the 80th percentile in a baseline random forest model

Out of the demographic features, age proved to be the most significant, which aligns with the body of research showing the risk of neurological disorders increasing with age. Among the voice analysis, the spectral centroid mean was the best predictor of Parkinson's disease. This feature refers to the center of mass of a voice signal and often corresponds to how sharp or muffled the sound is, corresponding to the vocal changes observed in PD. For the gait features, the root mean square and standard deviation of acceleration in the z direction had the highest feature importance. The root mean square, corresponding to the magnitude of acceleration, shows how forcefully the participants moved up and down when walking, while the standard deviation shows the variability of vertical acceleration, corresponding to shaky motion, a hallmark of Parkinson's disease.

For the raw tapping information, the number of left taps, right taps and total taps quantify how many times the participants were able to tap their screen in 20 seconds, corresponding to their dexterity. Meanwhile, tapping consistency refers to the standard deviation of the time between taps, showing whether they kept a consistent pace throughout the test. In addition to the raw tapping information, the tapping acceleration measurements from the participants' smartphones were also significant. Among these features, the root mean square, or magnitude of acceleration, as well as the standard deviations of accelerations in each direction, which captures abrupt movements characteristic of PD. Then, in the frequency domain, the average frequency and spectral centroid capture the smoothness and consistency of tapping acceleration. Finally, the spectral spread of tapping acceleration serves as an additional indicator of the erraticness of the tapping motion, detecting tremors.

For input into the proposed qSVM model, which is highly sensitive to feature ordering and magnitude [22], we multiplied each feature by its importance. We sorted them accordingly to emphasize the significance of higher-performing features. We then scaled all features by a factor of 10 to ensure that all features had a magnitude close to 1, enhancing the model's ability to process the data effectively.

*Model Architecture:*

We explored quantum computing, specifically Quantum Support Vector Machines (qSVMs) due to their potential in accurately classifying high-dimensional datasets similar to those we work with (mPower). Quantum SVM (qSVM) models can access high-dimensional quantum Hilbert spaces, allowing them to encode complex relationships more effectively than standard classification models [23]. For the mPower data, which includes diverse and complex biomarkers like voice, gait, tapping, and demographic features, qSVMs can leverage quantum feature mapping to capture subtle, non-linear interactions between these heterogeneous variables. Researchers have increasingly applied qSVMs in clinical diagnosis [24]. However, quantum computing in the current noisy-intermediate scale quantum (NISQ) era remains costly, time-intensive, and error-prone [25]. To address these challenges, our study introduces a quantum-inspired kernel architecture that we can easily simulate on classical hardware, while still outperforming traditional models. However, unlike many current qSVM kernels, our model does not rely on entanglement, which is challenging to simulate classically; instead, it uses dynamic angle embedding.

*Evaluation and Comparative Analysis:*

Once we constructed the custom qSVM architecture, we then trained the model on our dataframe of 194 samples. Afterward, we compared the accuracy, ROC/AUC score, and precision to current state-of-the-art models in the field to demonstrate the viability of our approach. These benchmarks included architectures that have been explored for medical applications in the past, such as neural networks, SVM and qSVM models, and random forest. The comparative results are displayed below:

| Model | Accuracy | ROC/AUC | F1 Score | Precision | Recall |
|---|---|---|---|---|---|
| **Proposed** | **0.90** | **0.98** | **0.90** | 0.90 | **0.90** |
| **Classical ML** | | | | | |
| *Linear SVM* | 0.77 | 0.96 | 0.79 | 0.82 | 0.77 |
| *Polynomic SVM* | 0.70 | 0.92 | 0.76 | 0.84 | 0.70 |
| *RBF SVM* | 0.70 | 0.89 | 0.74 | 0.78 | 0.70 |
| *Random Forest* | 0.60 | 0.94 | 0.69 | 0.81 | 0.60 |
| *Gradient Boost* | 0.63 | 0.89 | 0.71 | 0.82 | 0.63 |
| *Naive Bayes* | 0.63 | 0.92 | 0.68 | 0.75 | 0.63 |
| *Logistic Regression* | 0.73 | 0.97 | 0.79 | 0.85 | 0.73 |
| *KNN* [26] | 0.67 | 0.94 | 0.72 | 0.77 | 0.67 |
| *CNN* [18] | 0.60 | 0.67 | 0.60 | 0.60 | 0.60 |
| *FNN* [25] | 0.67 | 0.94 | 0.72 | 0.77 | 0.67 |
| *DNN* [26] | 0.80 | 0.93 | 0.82 | 0.93 | 0.74 |
| **Quantum SVM** | | | | | |
| *Z Feature Map* | 0.87 | 0.86 | 0.87 | 0.87 | 0.87 |
| *ZZ Feature Map* | 0.67 | 0.68 | 0.53 | 0.44 | 0.67 |

Table 1: Performance of various models across accuracy, ROC/AUC, precision and recall

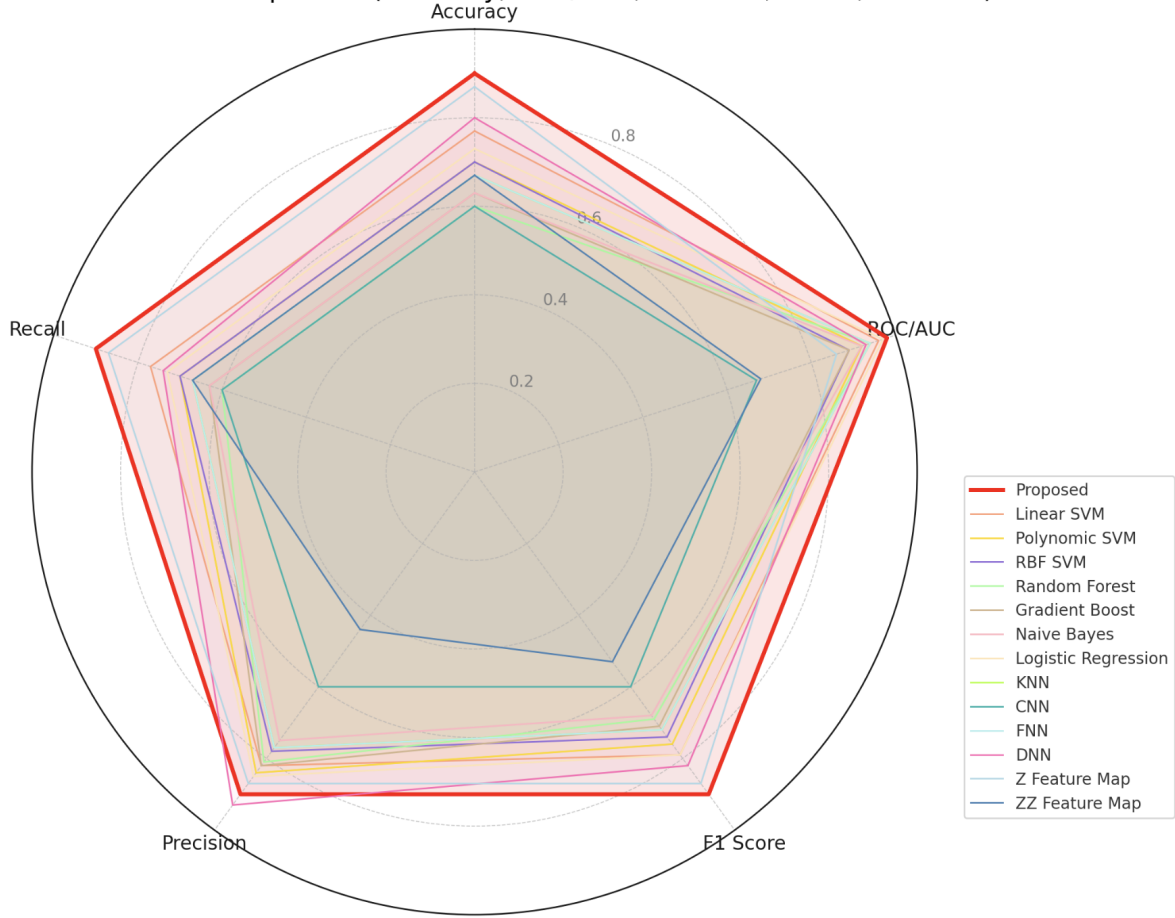

Figure 2: Radar chart for proposed model and benchmark comparison across accuracy, precision, recall, F1 score and ROC/AUC score

Due to the large number of features and the data's high complexity, models that incorporated strong overfitting protections generally performed better (Table 1). Among the classical algorithms, logistic regression and the linear SVM demonstrated the best performance. By contrast, complex neural networks tended to overfit, reducing their accuracy. Among alternative qSVMs, the entanglement-heavy ZZ feature map performed poorly, likely because classical simulators cannot emulate entanglement accurately. The Z feature map without entanglement performed better but lacked the complexity necessary to capture the full dimensionality of the data, as reflected in its lower metrics. The proposed kernel, by using quantum rotation gates to encode features into a complex quantum state without requiring entanglement, achieved the highest performance across accuracy, ROC/AUC score and F1 score (Figure 2), emphasizing its potential as a baseline for future clinical applications.

**Discussion**

In this study, we rely on a multimodal framework for prediction, using voice recordings with a universal vowel sound "ahh", gait indications from phone-detected acceleration data points, phone tapping count and acceleration, as well as demographic information. We demonstrated the efficiency and feasibility of our novel simulatable qSVM with a custom kernel involving quantum rotation gates, resulting in comparably superior performance metrics overall, and more productive testing procedures, without the need for a fault-tolerant quantum computer.

*Limitations*

Concerns could arise regarding the model's ability to generalize to new data. With only a train and test dataset, there is a possibility of the model overfitting to the validation set. Overfitting occurs when a model becomes overly complex, relying on the noise within the data rather than solely focusing on underlying patterns, causing the model to generalize poorly to real-world data [27]. In the future, this could be improved by conducting rigorous clinical trials, where the model is exposed to a wider range of patient populations, clinical settings, and environmental factors, testing the model on real-world data. This would serve as a cross-validation of model performance. Moreover, data collection conditions similar to the mPower study should be replicated to ensure optimal results.

*Ethics*

All data used was made available by the mPower public researcher portal, and our academic usage falls within the guidelines outlined in the research license. In a real-world application, we acknowledge that false positives, though rare, could have the potential to cause unnecessary stress. Therefore, we assert that this model is not a standalone diagnostic but rather a screening tool, to be considered in the larger context of overall health by medical professionals. Furthermore, our work is intended to serve as a baseline for further research and testing.

*Broader Impact & Future Work*

Our goal is to further research in the field of PD prediction, particularly multimodal early screening tools that combine input data from multiple biomarkers to improve classification accuracy. Such noninvasive approaches allow for greater accessibility of diagnosis while lowering costs compared to traditional methods. Additionally, the inclusion of multiple features not only improves model performance metrics, but also enables a more personalized prediction, leading to more tailored treatment plans further down the line. Moreover, we encourage future studies to explore simulatable qSVM kernels for a variety of real-world applications, potentially leading to new breakthroughs in various fields.

In summary, we propose a multimodal classifier for Parkinson's disease built on a diverse dataset representing the global population. This model leverages a quantum Support Vector Machine (SVM) to make predictions on the basis of tapping measurements, gait tracking, voice recordings, and demographic information. For optimized resource usage, the underlying quantum kernel used relies on rotation gates rather than entanglement operations, allowing the model to be simulated classically. This framework harnesses the potential of quantum algorithms while removing the need for expensive quantum hardware, improving access to PD prediction for millions worldwide.

**Methods**

*Gait Data Processing*

Although the mPower study included gait tracking measurements from a variety of different devices, including pedometers and accelerometers, we decided to focus on the smartphone measurements to preserve the accessibility and broad applicability of our model. The smartphone measurement file was composed of time series data tracking the x, y, and z acceleration of the device at regular intervals while the participant was walking. In order to interpret the data, we extracted the root mean square of the acceleration in the x, y, and z directions as well as that of the total signal, a metric commonly used to capture magnitude when describing time-varying quantities [28]. In addition, we included the standard deviation of acceleration in each direction to capture the variability over time. Next, we converted the total magnitude of acceleration into the frequency domain using a fast Fourier transform. In the frequency domain, we extracted the dominant frequency, mean frequency, spectral centroid, and spectral spread, corresponding to signal features that have the potential to capture the characteristic erratic movements of individuals affected by Parkinson's disease [29].

*Tapping feature extraction*

In the tapping test, mPower participants used a smartphone app to alternate tapping a left button and a right button as many times as they could in 20 seconds. From this, two types of measurements were recorded. First, each left tap and right tap were recorded, along with their corresponding timestamps. From this information, we extracted the number of taps on the left button, right button, and the total number of taps in 20 seconds, which shows the dexterity of the participants. Next, we extracted the number of repeated taps on the same button, which shows the participant's ability to alternate between buttons effectively. Finally, we extracted the consistency of their taps, which we measured by computing the standard deviation of the time between taps, corresponding to how consistent their tapping speed was.

The second type of measurement recorded was the acceleration of the smartphone in the x, y, and z directions throughout this 20 second interval. Similarly to gait, we extracted the root mean square and standard deviation of acceleration in the x, y and z directions, as well as for the total signal. In the frequency domain, we extracted the dominant and mean frequencies, in addition to the spectral centroid and spectral spread. Through all of the acceleration features, we hoped to represent the characteristic shakiness and tremors associated with PD [30].

*Vocal feature extraction*

While gait and demographic, and tapping data were expressed numerically, the researcher portal provided vocal data in the form of 10 second recordings of the vowel sound "ahhh". To quantify the differences in voice, specific vocal characteristics were considered. Variations in volume and pitch have been known to be associated with PD [31]. As a result, their means and standard deviations were added to the dataframe. Furthermore, signal characteristics also capture variations in voice, so the means and standard deviations of the zero-crossing rate, root mean square, spectral centroid, spectral bandwidth, and spectral rolloff were included as well. Research has described Parkinson's affected voices as having an airy or breathy quality [32], and direct quantification of this feature was performed using the Acoustic Breathiness Index proposed by [33]. This index was added to the feature list, along with its inputs of harmonics to noise ratio, cepstral peak prominence, power spectral density, harmonic difference, glottal to noise excitation ratio, high-frequency noise occurrence at 6000 Hz, and shimmer in decibels. Finally, Mel Frequency Cepstral Coefficients were extracted for each recording, with Principal Component Analysis being applied to reduce the number of MFCC features to ten for computational efficiency.

*Demographic features*

In addition to professional diagnosis, various information about the participants was provided. Excluding the hardware specifications, participant data mainly consisted of medical history. However, most of these measurements, such as past surgery, were not available for the majority of participants, hindering their efficacy. Furthermore, we excluded race to mitigate bias in detection. As such, age and smoking history were included due to a strong correlation with PD prevalence [34][35], as well as gender because of its effect on vocal characteristics [36].

*SVM Classifiers*

Support vector machine (SVM) models are widely used in binary classification problems due to their flexibility across dataset sizes and resistance to overfitting [37]. By plotting each datapoint in high-dimensional space, the kernel function of an SVM seeks to define a hyperplane boundary between each class. In this way, the model is able to classify new data by plotting it on this same plane and seeing which side of the boundary it falls on. These boundaries, or kernels, can be linear, polynomial, or a radial basis function (RBF).

*qSVM Model Architecture*

Quantum SVMs, or qSVMs, are being researched for their ability to improve classification by accessing high-dimensional Hilbert space. However, with current limitations on quantum hardware, quantum-inspired kernels that can be simulated on classical computers are more promising for real-world applications [38].

Currently, to construct quantum SVM models, Qiskit offers the Fidelity Kernel constructor, which mainly runs on the ZZ Feature Map [21]. This feature map has been known to achieve excellent performance on resource-intensive quantum hardware [22]. However, due to the feature map's inherent complexity and reliance on entanglement, which can be difficult to simulate classically, this kernel performs poorly in classical state vector simulations, as supported by Simoes et al [39].

Past papers, such as Kariya et al. and Suzuki et al., have explored the use of specific rotational gates to encode data characteristics into a complex quantum state, in conjunction with or as an alternative to entanglement [40, 41]. This study proposes the usage of Pennylane's provided Angle Embedding function [42], which encodes numerical data into rotation angles. This approach enhances the simplicity of the kernel construction by mitigating resource-intensive matrix operations or entanglement.

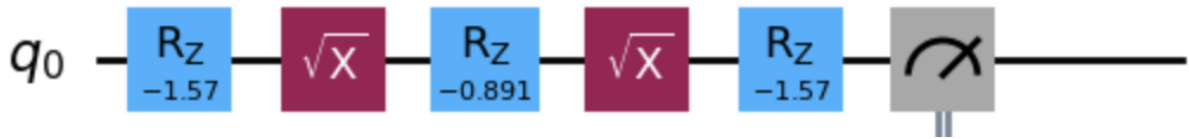

Figure 3: Example qubit encoding of one feature of the data

SVM kernels work by computing the similarity between two data points $x_1$ and $x_2$. To do this through quantum operations, each qubit represents one feature of the data. For each feature, the proposed model first performs a Y axis rotation of the $x_1$ value, followed by a Y axis rotation of $-x_2$. If $x_1$ and $x_2$ are similar, the resulting measurement would yield a value close to the qubit's initial state of 0 since the rotations would nearly cancel out. In this way, by measuring the magnitude of the final qubit state, the kernel computes the overlap between $x_1$ and $x_2$.

For ease of computation, the kernel replaces two rotations of $R_Y(x_1)$ and $R_Y(-x_2)$ with a single rotation of $R_Y(x_1 - x_2)$. As shown in Figure 3, Qiskit breaks down $R_Y$ operations into $R_Z$ and $R_X$ components, using quantum mechanical identities. Since $R_X(\frac{\pi}{2})$ is equivalent to $\sqrt{x}$, the kernel can be programmed as $R_Z(-\frac{\pi}{2})$, $\sqrt{x}$, $R_Z(x_1 - x_2)$, $\sqrt{x}$ and $R_Z(-\frac{\pi}{2})$, as shown.

After each qubit is measured, the values are aggregated through a weighted sum of each measurement multiplied by the random forest feature importance of the corresponding feature, transformed using a softmax function to emphasize the contribution of features with high predictive power. This computation is repeated on every pair of datapoints in the training set to construct the kernel matrix, which is then passed into a classical SVM for evaluation.

*Benchmark Models*

All benchmark models were chosen from either standard machine learning options or models used by prior researchers in this field [24] [18] [25] [26]. For most models, the same training and testing sets as the proposed model were used to ensure a fair comparison. However, for the alternative qSVM kernels of the Z and ZZ feature map, the full dataset was too resource-intensive to run. So, we chose to extract metrics based on a subset of the dataset including the first 30 train and 15 test samples.

**Data Availability**

All training data was made available by the [mPower Public Research Portal](mPower Public Research Portal) [20], which aligns with IRB approvals and permits academic usage through its user guidelines.